\documentclass[aps,prl,twocolumn,amsmath,amssymb,amsfonts,nofootinbib,long,floatfix]{revtex4}
\usepackage{graphicx,epsfig,latexsym,bm}

\begin{document}

\title{Dimensional analysis of two-dimensional turbulence}

\author{Leonardo Campanelli$^{1}$}
\email{leonardo.s.campanelli@gmail.com}

\affiliation{$^1$All Saints University, Asudom Academy of Science, 5145 Steeles Ave., Toronto (ON), Canada}

\date{\today}


\begin{abstract}
We study the scaling properties of two-dimensional turbulence using
dimensional analysis. In particular, we consider the energy spectrum both at large and small
scales and in the ``inertial ranges'' for the cases of freely decaying and forced turbulence.
We also investigate the properties of an ``energy condensate'' at large scales in spatially
finite systems. Finally, an analysis of a possible inverse cascade in freely decaying
turbulence is presented.
\end{abstract}


\maketitle


\section{I. Introduction}

Dimensional analysis is a powerful tool in the analysis of turbulence
when combined with the results of direct numerical simulations, laboratory experiments, and
approximate analytical solutions. Indeed, the aim of this paper is to analyze
two-dimensional hydrodynamic (2HD) turbulence using only dimensional analysis under appropriate
hypothesis derived from existing theoretical and experimental studies
(for a recent review on two-dimensional turbulence, see~\cite{Boffetta}).

The most important quantity in locally homogeneous and isotropic
turbulence is the so-called energy spectrum $E(k,t)$, which defines the energy
contained in a given (Fourier) velocity mode
of wavenumber $k$ at the time $t$. Integrating over $k$, one obtains the energy
$E(t) = \int_0^{\infty} \! dk E(k,t)$,
which is one of the two invariants in 2HD turbulence in the limit of vanishing (kinematic)
viscosity $\nu$, the other one being
the so-called enstrophy $\Omega(t) = \int_0^{\infty} \! dk k^2 E(k,t)$.

The object of this paper, indeed, is the study of the energy spectrum both in the
case of freely decaying turbulence (no external sources of energy and enstrophy)
and when energy and enstrophy are injected at a given length scale (forced turbulence).

\section{II. Homogeneous and isotropic turbulence}

The most simple and successful mathematical model of turbulence is
based on the hypothesis, introduced by Taylor long time ago~\cite{Taylor},
of ``local'' homogeneity and isotropy.

Although an even simpler model of turbulence would be one in which the velocity field
is ``globally'' homogeneous and isotropic, independent on the initial conditions and external
systems, it is of little interest since it cannot
be realized in nature. In that case, indeed, the field would be infinite and unbounded,
with no mean velocity and no mean gradients, and would simply decay in time under the
action of the viscous term. Nevertheless, for reasons that will be apparent later,
we start our analysis by considering such an ideal case and in so doing we
follow Manton and Luxton~\cite{Manton} who considered the case of three-dimensional
turbulence.

On dimensional grounds, the (kinetic) energy spectrum $E(k,t)$ has the form
\begin{equation}
\label{1}
E(k,t) = \nu^{3/2} t^{-1/2} \psi(k \sqrt{\nu t}),
\end{equation}
where $\psi$ is an arbitrary function of its argument.
The only scale in the model is the dissipation length $L_{diss}(t) = \sqrt{\nu t}$, to
which it corresponds the wavenumber $k_{diss}(t) = 1/L_{diss}$.

At small scales, $k \gg k_{diss}$, the 2HD equations
can be exactly solved if one assumes that the rate of energy transfer due to
nonlinear interactions is completely negligible. 
\footnote{The energy spectrum changes in time as $\partial E(k,t)/ \partial t = T(k,t) - \nu k^2 E(k,t)$,
where $T(k,t)$ is the energy change caused by nonlinear interactions~\cite{Boffetta}.
When this term is negligible with respect to the viscous term, the above equation
has the solution given below.}
In this case, we obtain $E(k,t) = f(k) e^{-\nu k^2 t}$, with $f(k)$
being an arbitrary function. Comparing with Eq.~(\ref{1}), we find
\begin{equation}
\label{2}
E(k,t) = c_{\infty} \nu^2 k e^{-\nu k^2 t}, ~~~ k \gg k_{diss},
\end{equation}
which corresponds to $\psi(x) = c_{\infty} x e^{-x^2}$, with
$c_{\infty}$ being a dimensionless constant.

We assume that the ``scaling function'' $\psi$ is analytic. The Taylor expansion at large scales
gives
\begin{equation}
\label{3}
E(k,t) = c_0 + c_1 k + c_2 k^2 + ..., ~~~ k \ll k_{diss},
\end{equation}
where $c_n(t) = \nu^{(n+3)/2} t^{(n-1)/2} \psi^{(n)}(0)/n!$.
The energy spectrum at $k \rightarrow 0$, which describes the properties of the
largest eddies of turbulence, is stationary only if $\psi(x) = C_1 x$,
where $C_1$ is a dimensionless constant. In this case, we have
$E(k,t) = C_1 \nu^2 k$.
The time-independence of the energy spectrum at small wavenumbers will be refereed to as
the {\it permanence of large-scale eddies} of turbulence.

\section{III. Freely decaying turbulence}

A physical homogeneous and isotropic field not interacting with external systems
is nevertheless dependent upon the initial conditions which
introduces a mean velocity $u_i$ and a length scale $L_i$. This is the case, for example,
of turbulence behind a grid (with $u_i$ being the velocity field at the grid
and $L_i$ being the spacing of the bars of the grid perpendicular to the direction of the flow).
\footnote{In numerical simulations, it is usually assumed that the initial velocity field is
homogeneous and isotropic. In this case, the mean value of the velocity is zero and $u_i$ in the
following equations must be regarded as a r.m.s value of the field. More generally,
$L_i$ and $u_i$ can be considered as typical scale values at the onset of fully developed
turbulence.}

Therefore, in freely decaying turbulence, we can introduce two new dimensionless
quantities~\cite{Manton},
\begin{equation}
\label{5}
M = \frac{u_i t}{L_i} \,, ~~~ R_i = \frac{u_i L_i}{\nu} \, .
\end{equation}
The first can be view as a dimensionless length scale and can be also rewritten as
$M = t/\tau_i$, where $\tau_i = L_i/u_i$ is the so-called initial turnover time,
while the second is the initial Reynolds number and can be regarded as a dimensionless
velocity scale.

Equation~(\ref{1}) must be then replaced by
\begin{equation}
\label{6}
E(k,t) = \nu^{3/2} t^{-1/2} \psi(k \sqrt{\nu t},M,R_i).
\end{equation}
(The use of the same symbol for the scaling function $\psi$, here
and in the following, should not cause confusion.)
In this case, there are two length scales: the dissipation length $L_{diss}$ and
the initial scale $L_i$, which we assume to be much greater than
$L_{diss}$, $L_i \gg L_{diss}$. The wavenumber corresponding to $L_i$ is
$k_i = 1/L_i$.

These two length scales define three ranges: ($i$) the large-eddies range,
$k \ll k_i$, ($ii$) the energy inertial range,
$k_i \ll k \ll k_{diss}$, and the dissipation range, $k \gg k_{diss}$ (see Fig.~$1a$).

{\it Dissipation range.} -- One would expect that in the dissipation range the energy
spectrum evolves as in Eq.~(\ref{2}).
However, Tatsumi and Yanase~\cite{Tatsumi}, using the ``modified zero
fourth-order cumulant approximation'',
found an analytic expression of the energy spectrum in the dissipation
range which corresponds to
\begin{equation}
\label{6a}
\psi(x,y,z) = 384 b^{3/2} x^{5/2} e^{-bx},
\end{equation}
where $b$ is a dimensionless constant. The same authors found, by
direct numerical integration of 2HD equations,
that $\psi(x,y,z) \propto e^{-bx^s}$ with $s$ in the range
$[1.3,1.4]$. The discrepancy with Eq.~(\ref{6a}) indicates, according to the authors,
that either the asymptotic behaviour $s = 1$ is realized beyond the numerical coverage or
the numerical results are not accurate
enough. In either case, the asymptotic form of the spectrum is
different from the purely viscous spectrum $\psi(x,y,z) \propto x e^{-x^2}$.


\begin{figure*}[t]
\begin{center}
\includegraphics[clip,width=0.48\textwidth]{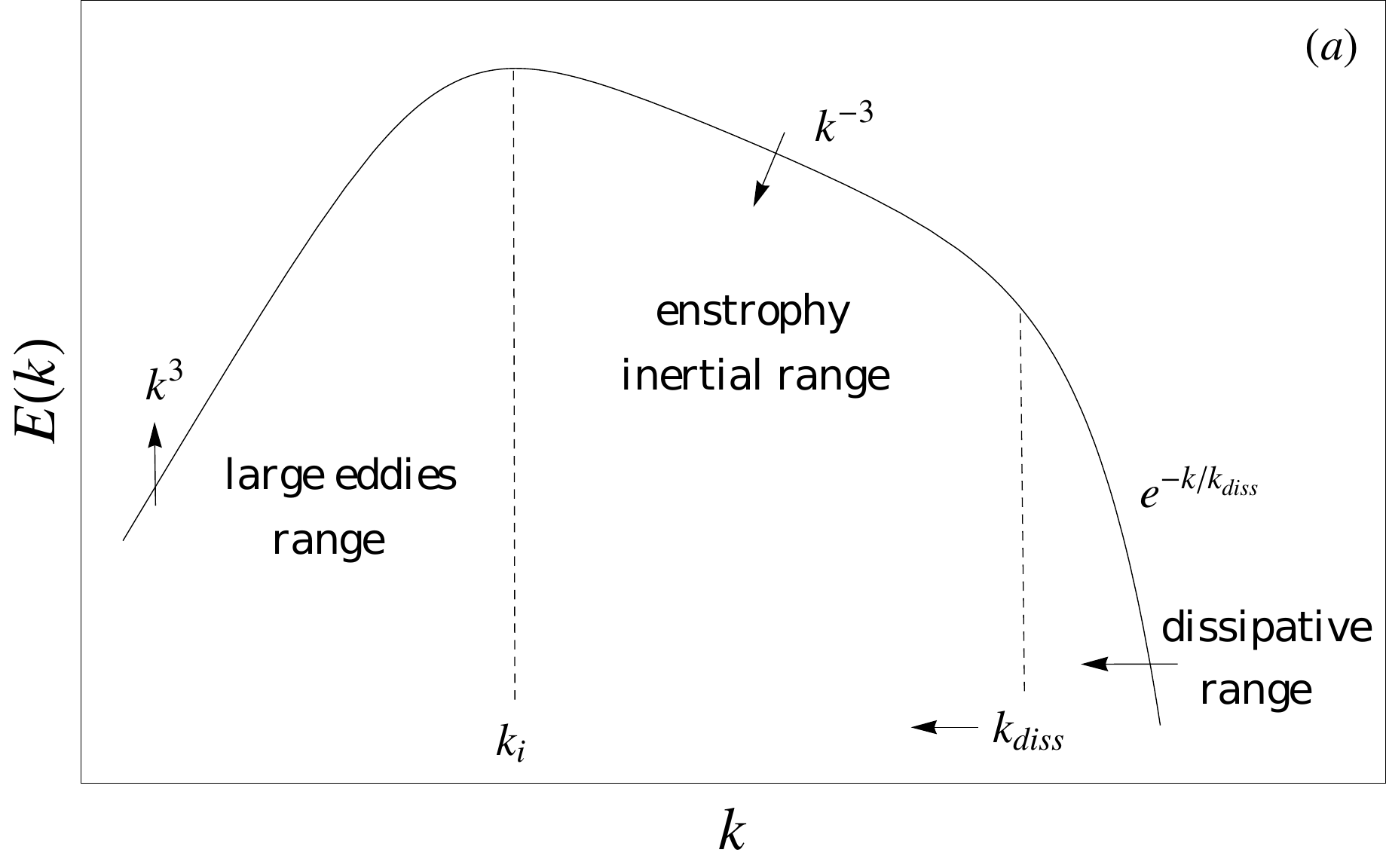}
\hspace{0.3cm}
\includegraphics[clip,width=0.48\textwidth]{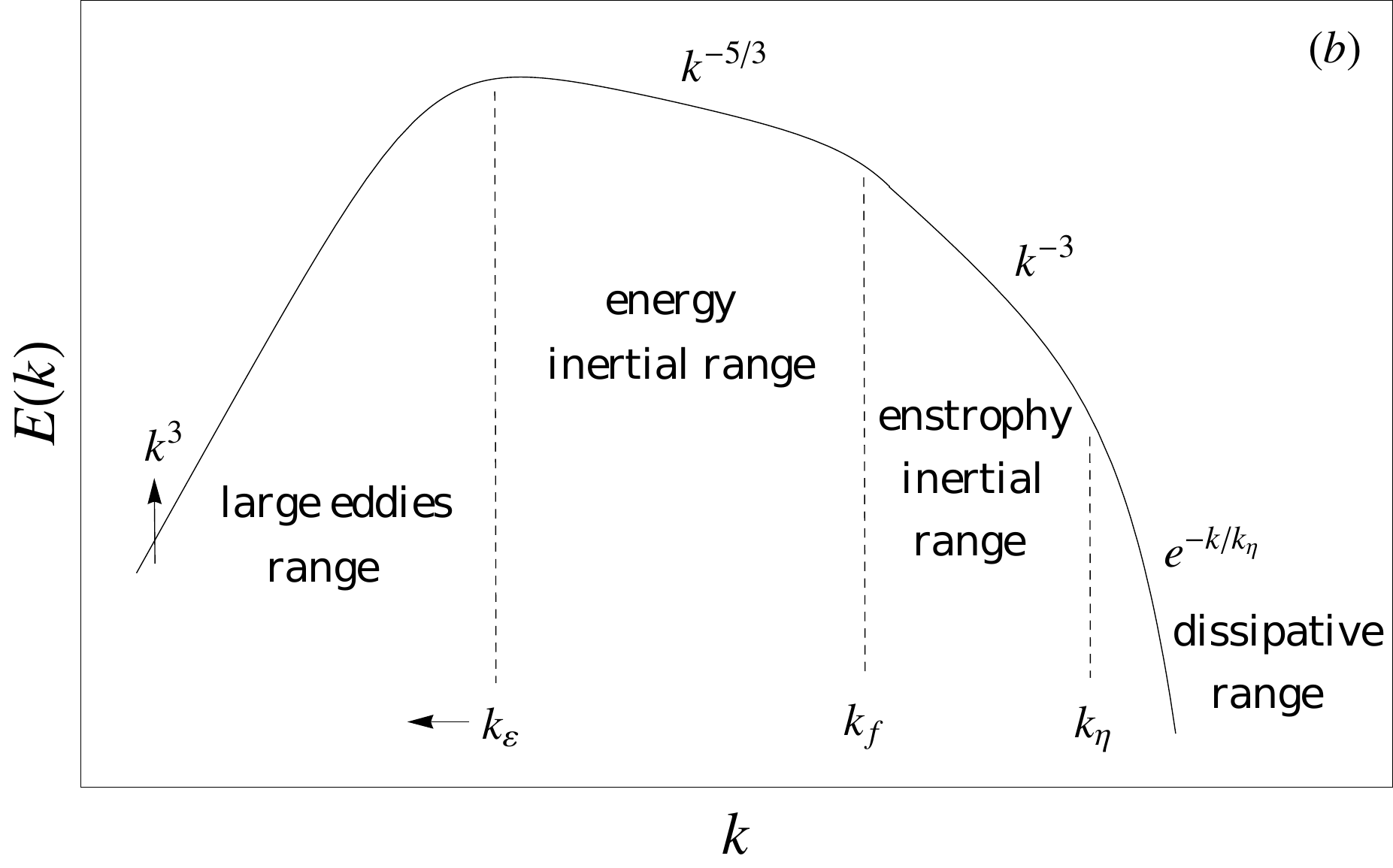}


\includegraphics[clip,width=0.48\textwidth]{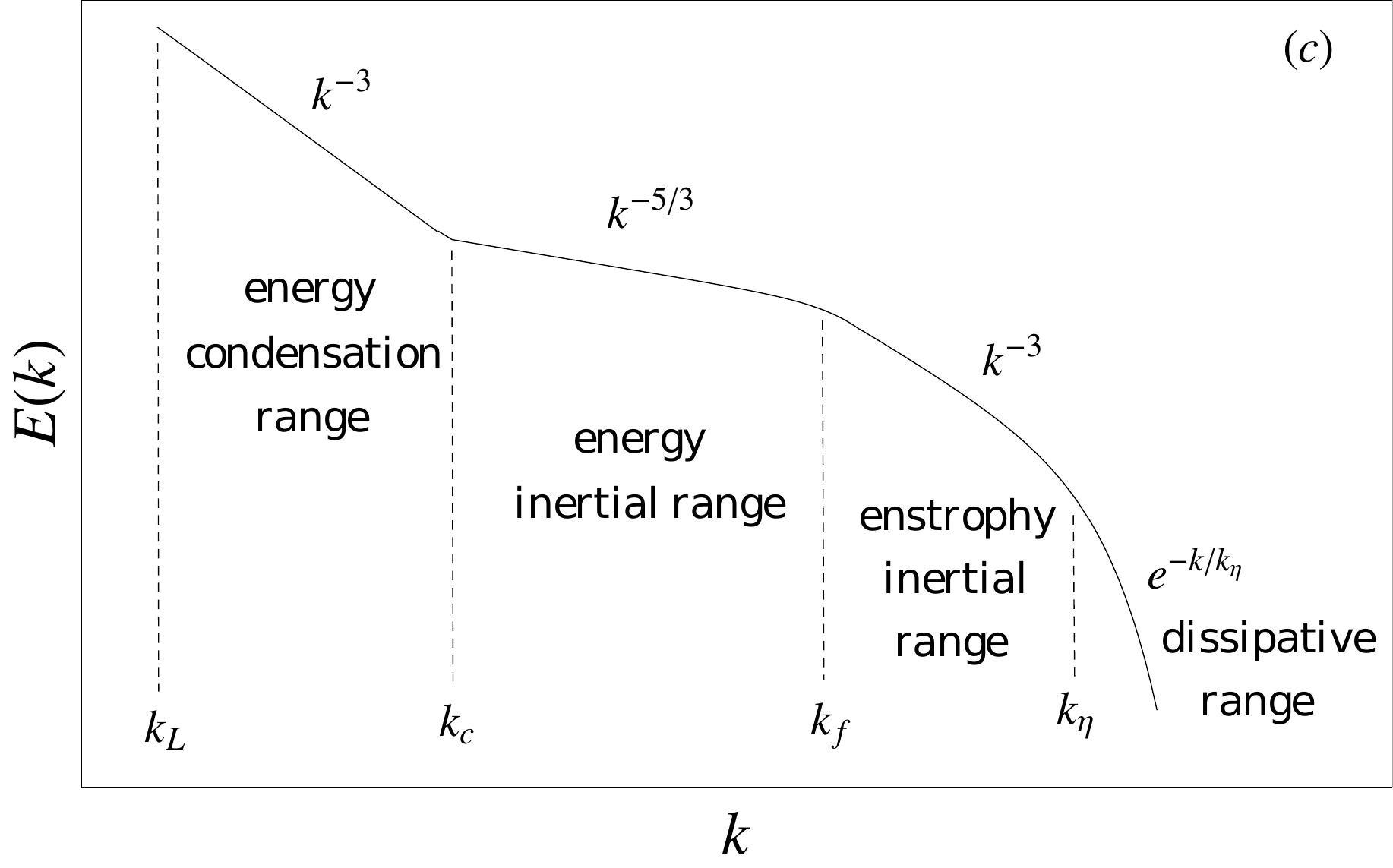}
\hspace{0.3cm}
\includegraphics[clip,width=0.48\textwidth]{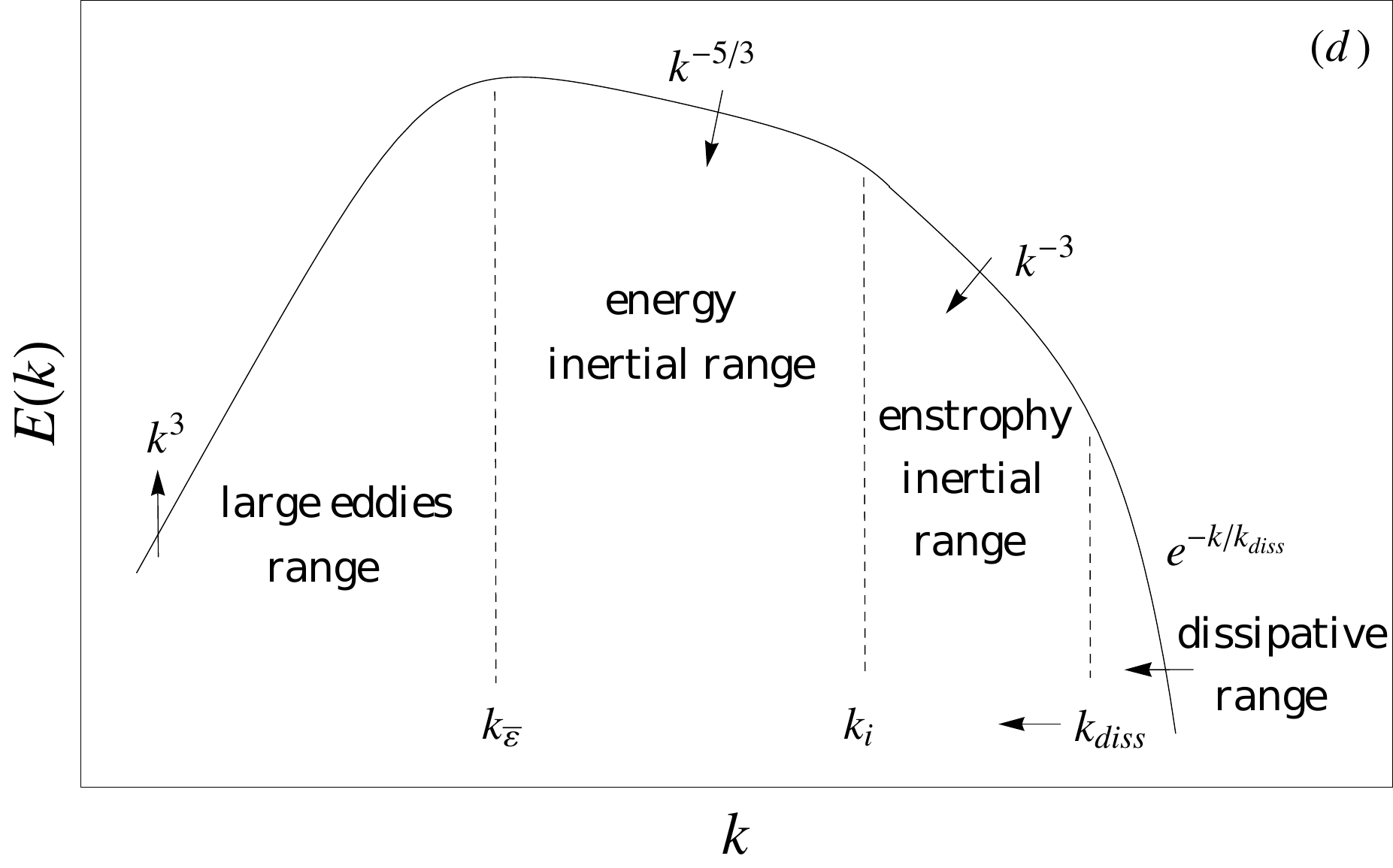}
\caption{Picture (log-log scale) of the expected kinetic spectrum in freely decaying ($a$)
and forced ($b$) two-dimensional turbulence as a function of the wavenumber.
Full spectrum in the presence of energy condensation ($c$) and
inverse cascade in freely decaying turbulence ($d$). Arrows indicate displacements in time.
In ($a$), ($b$), and ($c$), a time-dependent Saffman spectrum is shown at large scales
($k \rightarrow 0$).}
\end{center}
\end{figure*}


{\it Large-eddies range: the Saffman integral.} -- Let us first assume a
permanence of large-scale eddies. The energy spectrum for $k \rightarrow 0$, then,
does not depend on time and this happens only if $\psi(x,y,z) = x \psi(xy^{-1/2},z)$ or
\begin{equation}
\label{7}
E(k,t) = \nu^2 k \psi(kL_i/\sqrt{R_i},R_i).
\end{equation}
For $k \rightarrow 0$, we then have $E(k,t) = \nu^2 k \psi(0,R_i)$
(assuming that the limit exists and is different from zero). Because the spectrum does
not depend on the dissipation parameter $\nu$ for large Reynolds numbers, we must have
$\psi(0,R_i) = \bar{c}_0 R_i^2$, where $\bar{c}_0$ is a dimensionless constant.
Accordingly,
\begin{equation}
\label{8}
E(k,t) = \mathcal{L} k, ~~~ k \rightarrow 0,
\end{equation}
where $\mathcal{L} = \bar{c}_0 u_i^2 L_i^2$. A linear, infrared spectrum
was predicted by Saffman long time ago~\cite{Saffman}. The quantity $\mathcal{L}$ is
known as the two-dimensional Saffman integral~\cite{DavidsonBook} and its
time-independence is a consequence of the conservation of linear momentum~\cite{Davidson}.

{\it Large-eddies range: the Loitsyansky integral.} -- The principle of permanence of
large-scale eddies is well confirmed in three-dimensional turbulence. In direct numerical
simulations of two-dimensional turbulence, instead, an increase in time of the energy
spectrum at large scales is usually observed. If this is due to the the finiteness of the
simulation box or is a genuine effect, that is still unclear. Recently enough, however,
Davidson~\cite{Davidson} has shown, under plausible hypotheses, that the infrared spectrum
has the form
\begin{equation}
\label{8a}
E(k,t) = I(t) k^3, ~~~ k \rightarrow 0,
\end{equation}
where $I(t)$ is the so-called Loitsyansky integral in two dimensions.
Following Davidson and assuming $I(t) \propto t^{5/2}$, is then
easy to see that dimensional analysis gives
\begin{equation}
\label{8b}
I(t) = C_0 L_i^{-3/2} u_i^{3/2} t^{5/2},
\end{equation}
where $C_0$ is a dimensionless constant and we assumed that the spectrum does not depend
on the dissipation parameter. Equation~(\ref{8b}) corresponds to take
$\psi(x,y,z) = C_0 x^3 y^{3/2} z^3$.
\footnote{Equations~(\ref{8}) and (\ref{8b}) represent the first two terms of a Taylor
expansion, $E(k,t) = \mathcal{L} k + I(t) k^3 + \mathcal{O}(k^5)$. A Saffman spectrum
can be realized only when $I(t)$ is identically zero. If this is not the case, the
linear part of the spectrum is progressively ``overshadowed'' by the increasingly
strong cubic part~\cite{Davidson}.}

{\it Energy inertial range.} -- By definition, an inertial range is a range where
dissipation is negligible and the dynamics is independent on the initial conditions.
In this range, then, the energy spectrum is given by Eq.~(\ref{1}),
namely $\psi$ does not depend on $M$ and $R_i$. Independence on $\nu$ implies
$\psi(x) = c' x^{-3}$, where $c'$ is a dimensionless constant. Accordingly,
the only possible spectrum in this range is
\begin{equation}
\label{11}
E(k,t) = c' t^{-2} k^{-3}, ~~~ k_i \ll k \ll k_{diss}.
\end{equation}
In forced turbulence, a time-independent spectrum of the form $E(k,t) \propto k^{-3}$
is known as Batchelor spectrum (see next section).

\section{IV. Forced turbulence}

In forced turbulence, energy and enstrophy are injected at constant rates
at the forcing scale $L_f$, to which it correspond the wavenumber $k_f = 1/L_f$.
The dynamics is very well understood in this case.
Indeed, as first discussed by Kraichnan~\cite{Kraichnan} in 1967, there are two
inertial ranges that, starting from the forcing scale, extend to smaller and larger
scales, respectively.

{\it Enstrophy inertial range.} -- In this range, that extend from the forcing
scale down to the dissipation scale, enstrophy is transferred from larger to
smaller scales (direct cascade of enstrophy). Here, the dynamics is ruled by the
(constant) enstrophy flux $\eta$ instead of $t$~\cite{Batchelor}:
\begin{equation}
\label{12}
t \rightarrow \eta^{-1/3}.
\end{equation}
Substituting in Eq.~(\ref{1}), and imposing that the spectrum
does not depend on $\nu$, we find $\psi(x) = C' x^{-3}$,
where $C'$ is a dimensionless constant. This corresponds to
the Batchelor spectrum~\cite{Batchelor}
\begin{equation}
\label{12a}
E(k,t) = C' \eta^{2/3} k^{-3}, ~~~ k_f \ll k \ll k_\eta.
\end{equation}
Here,
\begin{equation}
\label{13}
k_\eta = \eta^{1/6} \nu^{-1/2}
\end{equation}
is the wavenumber corresponding to the enstrophy dissipation scale
$L_\eta = 1/k_\eta$ that is obtained from the viscous dissipation scale
$L_{diss}$ by means of the substitution~(\ref{12}).

{\it Energy inertial range.} -- In this range, that extends from the forcing
scale to larger scales, energy is transferred from smaller to larger scales
(inverse cascade of energy). The dynamics, this time, is ruled by the (constant)
energy flux $\varepsilon$~\cite{Kolmogorov}:
\begin{equation}
\label{13a}
t \rightarrow \nu^{1/2} \varepsilon^{-1/2}.
\end{equation}
Substituting in Eq.~(\ref{1}), and imposing independence on viscosity,
we obtain $\psi(x) = C x^{-5/3}$, to wit the famous Kolmogorov spectrum~\cite{Kolmogorov}
\begin{equation}
\label{14}
E(k,t) = C \varepsilon^{2/3} k^{-5/3}, ~~~ k_\varepsilon \ll k \ll k_f,
\end{equation}
where $C$ is a dimensionless constant. Here,
\begin{equation}
\label{15}
k_\varepsilon = \varepsilon^{-1/2} t^{-3/2}
\end{equation}
is the only wavenumber that can be constructed starting from $\varepsilon$ and $t$
(such a wavenumber should not depend either on the initial conditions or on the
dissipation parameter), and then represents the maximum wavenumber to which energy
is transferred from smaller scales. Note that the corresponding scale
$L_\varepsilon = 1/k_\varepsilon \propto t^{3/2}$ increases in time (see Section V).

{\it Large-eddies range.} -- The dynamics in the large-eddies range is the same as
in the case of freely decaying turbulence since the energy and enstrophy pumping at the
forcing scale does not affect modes with vanishing wavenumbers, $k \rightarrow 0$.

{\it Dissipation range.} -- As in the case of freely decaying turbulence,
the asymptotic form of the spectrum in the dissipation range is found to be
different from the purely viscous spectrum~\cite{Gotoh}.

In this range, the time is replaced by $\eta^{-1/3}$ [as in Eq.~(\ref{12})],
so that the dissipation wavenumber in forced turbulence, $k_\eta$,
takes the place of the dissipation wavenumber in freely decaying turbulence, $k_{diss}$.
Consequently, the spectrum has the form
\begin{equation}
\label{15a}
E(k,t) = \nu^{2} k_\eta \psi(k/k_\eta,M_\eta,R_i),
\end{equation}
with $M_\eta = u_i \eta^{-1/3} L_i^{-1}$, and we expect a scaling function
of the form $\psi(x,y,z) \propto x^{5/2} e^{-\beta x}$ [compare with Eq.~(\ref{6a})],
where $\beta$ is a constant.

Gotoh~\cite{Gotoh}, using the K\'{a}rm\'{a}n-Howarth-type equation, derived the spectrum
\begin{equation}
\label{15c}
E(k,t) \propto k^{-(3+\delta)/2} e^{-\alpha_2 k/k_\eta},
\end{equation}
with $\delta$ and $\alpha_2$ depending on the Reynolds number
(see~\cite{Gotoh} for details). The theoretical expectation~(\ref{15c})
was confirmed by Gotoh by a direct numerical simulations at moderate
Reynolds numbers. He also argued that, in the limit of large Reynolds numbers,
$\delta$ should vanish while $\alpha_2$ should approach a constant value
$\beta$, giving to the scaling function $\psi$ the universal form
\begin{equation}
\label{15d}
\psi(x,y,z) \propto x^{-3/2} e^{-\beta x}.
\end{equation}
It is interesting to observe that the coefficients of the power-law, pre-exponential
factors for the case of freely decaying and forced turbulence are different.

In forced turbulence, then, there are four ranges: ($i$) the large-eddies range,
$k \ll k_\varepsilon$, ($ii$) the energy inertial range,
$k_\varepsilon \ll k \ll k_f$, the enstrophy inertial range,
$k_f \ll k \ll k_\eta$, and the dissipation range, $k \gg k_\eta$.
A sketch of the energy spectrum in these ranges is shown in Fig.~$1b$.

\section{V. Energy condensation}

{\it Forced turbulence.} -- Energy in forced turbulence is transferred from the
forcing scale to larger scales. If the system is contained, let us say, in a box of
linear dimension $L$, energy will tend to accumulate on such a scale, a
phenomenon known as ``energy condensation''~\cite{Boffetta}.
The transfer of energy to the scale $L$ takes place in a finite time, $t_{max}$,
that can be estimated by equating $k_\varepsilon$ in Eq.~(\ref{15}) to $k_L = 1/L$:
\begin{equation}
\label{16}
t_{max} = L^{3/2} \varepsilon^{-1/3}.
\end{equation}
Accordingly, after a time $t_{max}$ one would expect a Kolmogorov spectrum extending from
the forcing scale to $L$. However, both numerical simulations (see, e.g.,~\cite{Chertkov})
and laboratory experiments (see, e.g.,~\cite{Xia}) seem to indicate that this is not the case.
The spectrum substantially deviates from the Kolmogorov one, being consistent with a
$k^{-3}$ power-law behavior. In particular, a full spectrum of the form shown in Fig.~$1c$
has been observed in a recent simulation~\cite{Brandenburg},
\footnote{The inertial ranges in the numerical simulation of Ref.~\cite{Brandenburg}
are very narrow. Nevertheless, the $k^{-3}$ behaviour at the largest scales is very 
well resolved and extends on more than about one decade in wavenumber.}
with energy contained in the ``condensation range'' $k_L \leq k \leq k_c$ 
(an estimate of $k_c$ is given below).

The existence of the condensation range can be explained by dimensional analysis as follows.
Because energy in this range is transferred from the energy inertial range, the scaling
function $\psi$ should not depend on the initial conditions. However, the boundary effects
due to the finiteness of the space are important and introduce a new dimensionless length scale:
\begin{equation}
\label{16a}
M_L = \frac{u_i t}{L} \, .
\end{equation}
For fully developed turbulence, we expect that the energy spectrum near the maximum scale $L$
does not explicitly depend on the Reynolds number at that scale, $R_L = u_i L/\nu$.
Therefore, for the spectrum of the condensate, we can write
\begin{equation}
\label{17}
E(k,t) = \nu^{3/2} t^{-1/2} \psi(k \sqrt{\nu t},M_L).
\end{equation}
Under stationary conditions, we have $\psi(x,y) = x \psi(xy^{-1/2})$, which gives
\begin{equation}
\label{18}
E(k,t) = \nu^2 k \psi(k\sqrt{L\nu/u_i}).
\end{equation}
Independence on viscosity finally implies $\psi(x) \propto x^{-4}$, so that
\begin{equation}
\label{19}
E(k,t) = J_3 k^{-3}, ~~~ k_L \lesssim k \lesssim k_c,
\end{equation}
where $J_3 = c u_i^2/L^2$ and $c$ is a dimensionles constant.

The condensation scale $L_c = 1/k_c$ can be estimated by equating
Eqs.~(\ref{14}) and (\ref{19}) evaluated at $k = k_c$. We find
\begin{equation}
\label{20}
k_c = \hat{c} (u_i/L)^{3/2} \varepsilon^{-1/2},
\end{equation}
where $\hat{c} = (c/C)^{3/4}$.

{\it Freely decaying turbulence.} -- Davidson~\cite{Davidson} has shown
that in freely decaying turbulence a divergent spectrum of the form
\begin{equation}
\label{20a}
E(k,t) = J k^{-1}
\end{equation}
can be realized under particular initial conditions (in this case,
the spectrum at large scales is dominated by a sea of randomly located
monopole vortices). A non-null (time-independent) ``Davidson integral'' $J$
is associated to a net vorticity contained in a large two-dimensional
volume~\cite{Davidson}.

Needless to say, a divergent spectrum cannot be realized if the system is
spatially unbounded. For this reason, we consider a system confined in a
box of side $L$ and we require, in addition to the independence on time
and viscosity, that the spectrum is independent on the external scale $L$.
In this case, the energy remains finite even in the limit $L \rightarrow \infty$.

We start with Eq.~(\ref{6}) with $M$ and $R_i$ replaced by $M_L$ and $R_L$.
Time independence, then, gives Eq.~(\ref{7}) with $L_i$ and $R_i$ replaced by
$L$ and $R_L$. Independence on viscosity, instead, gives
$\psi(x,y) = y^2 \psi(xy^{1/2})$, or
\begin{equation}
\label{20b}
E(k,t) = u_i L^2 k \psi(kL).
\end{equation}
The only form of the spectrum which is independent on $L$ requires
$\psi(x) = \tilde{c}_0 x^{-2}$, where $\tilde{c}_0$ is a dimensionless
constant. This, in turns, gives Eq.~(\ref{20a}) with $J = \tilde{c}_0 u_i^2$.

\section{VI. Inverse cascade in freely decaying turbulence}

Mininni and Pouquet~\cite{Mininni} observed, for the first time,
an inverse cascade of energy in a series of direct numerical simulations
of freely decaying turbulence at moderate Reynolds numbers. 
They found that after a transient phase,
whose duration was of the order of few initial turnover times,
the energy present in the initial state and peaked at the scale $L_i$ was
transferred to larger scales. Moreover, they found that the spectrum in the
energy ``inertial'' range was well described by a $k^{-5/3}$ Kolmogorov-type
spectrum. However, the spectrum was not stationary, like in forced turbulence,
but a decay in time was observed. Such a decay is easy to understand. Indeed,
dimensional analysis shows that if the energy spectrum depends only on the
wavenumber $k$ and the energy flux $\overline{\varepsilon}$ (the energy per
unit time that is transferred to large scales), then it must takes on the
Kolmogorov form
\begin{equation}
\label{21}
E(k,t) =  \overline{C} \, \overline{\varepsilon}^{2/3} k^{-5/3},
\end{equation}
where $\overline{C}$ is a dimensionless constant. Since the energy flux must decreases
in time so does the energy spectrum. Indeed, since the energy and enstrophy
inertial ranges are adjacent, the spectra in Eqs.~(\ref{11}) and (\ref{21})
must match at $k = k_i$ at all times (as indeed observed in~\cite{Mininni}).
This gives
\begin{equation}
\label{22}
\overline{\varepsilon}(t) = \overline{c} \, L_i^2 t^{-3},
\end{equation}
where $\overline{c} = (c'/\overline{C})^{3/2}$. It is interesting to notice
that the energy flux depends only on $L_i$ but not on $u_i$.

The scale $L_{\overline{\varepsilon}}$ up to which energy is transferred from
the scale $L_i$ must approach a constant value after the transient phase since
there is not a constant injection of energy as in the case of forced turbulence.
Repeating the reasoning in Sec.~IV, the wavenumber $k_{\overline{\varepsilon}}$
corresponding to $L_{\overline{\varepsilon}}$ is
$k_{\overline{\varepsilon}} = \overline{\varepsilon}^{-1/2} t^{-3/2}$.
After inserting Eq.~(\ref{22}), it becomes
\begin{equation}
\label{23}
k_{\overline{\varepsilon}} = \overline{c}^{-1/2} k_i,
\end{equation}
namely a time-independent wavenumber, as expected.

A drawing of the energy spectrum in freely decaying turbulence with
inverse cascade of energy is shown in Fig.~$1d$.

Finally, let us find, for the sake of completeness, the form of the scaling
function in this case. The dynamics of inverse cascade depends on $M$ and $R_i$,
since the latter is initiated and sustained by the initial energy contained in
a range around the initial scale $L_i$. Thus, the appropriate form of the energy
spectrum is that in Eq.~(\ref{6}). Although the spectrum does not depend on $\nu$,
it depends on the initial conditions. For this reason, the range where energy is
transferred is not properly an ``inertial'' range. Combining Eqs.~(\ref{21}) and
(\ref{22}) we obtain $E(k,t) = c' L_i^{4/3} t^{-2} k^{-5/3}$, which corresponds to
a scaling function $\psi(x,y,z) = c' x^{-5/3} y^{-2/3} z^{2/3}$.

\section{VII. Conclusions}

Using dimensional analysis, and appropriate working hypotheses, we have obtained
the form of the energy spectrum in two-dimensional turbulence at large scales and
in the ``inertial ranges'', where the dynamics in independent on viscous dissipation
effects. We have re-obtained classical results in forced turbulence, such as the
Batchelor and Kolmogorov spectra in the enstrophy and energy inertial ranges,
respectively. When the external forcing is absent the Batchelor
spectrum decays in time, a result already known in the literature. 
Dimensional analysis, however, predicts a decays in time as fast as $t^{-2}$.
Also, dimensional analysis correctly predicts that the only stationary
(time-independent) spectrum independent on viscosity that describes the large-scale
eddies of turbulence is the Saffman spectrum, linear in the wavenumber.

When forcing is present, an accumulation of energy at scales as large as the physical
dimensions of a finite system is expected. Such a phenomenon of ``energy condensation'',
observed both in numerical simulation and laboratory experiments has, to our knowledge,
never been explained theoretically. Our dimensional analysis, indeed, predicts a 
spectrum at large scales of the form $k^{-3}$, in agreement with simulations and experiments. 
In freely decaying turbulence, and under particular initial conditions, an energy condensation 
can also occur. Dimensional analysis predicts in this case a $k^{-1}$ spectrum, 
in agreement with analytical calculations by Davidson.

Finally, we have discussed the dynamics of a possible inverse cascade in freely decaying
turbulence, as recently observed in numerical simulations by Mininni and Pouquet. In this
case, our results suggest that the energy flux, which drives the transfer of energy from the
initial length scale to larger scales, decays in time as $t^{-3}$.



\begin{thebibliography}{99}

\bibitem{Boffetta}      G.~Boffetta and R.~E.~Ecke, Annu.~Rev.~Fluid Mech. {\bf 44}, 427 (2012).

\bibitem{Taylor}        G.~I.~Taylor, Proc.~Roy.~Soc., Series A, Vol.~151, No.~873, 421 (1935).

\bibitem{Manton}        M.~J.~Manton and R.~E.~Luxton, ``Note on the Decay of Isotropic Turbulence'',
                        {\it Proceedings of the Third Australasian Conference on Hydraulics and Fluid Mechanics,}
                        Sidney, Australia (1968).

\bibitem{Tatsumi}       T.~Tatsumi and S.~Yanase, J.~Fluid Mech. {\bf 110}, 476-496 (1981).

\bibitem{Saffman}       P.~G.~Saffman,  
                        J.~Fluid Mech. {\bf 27}, 581 (1967).

\bibitem{DavidsonBook}  P.~A.~Davidson, {\it Turbulence},
                        (Oxford University Press, New York, 2004).

\bibitem{Davidson}      P.~A.~Davidson, J.~Fluid Mech. {\bf 580}, 431 (2007).

\bibitem{Kolmogorov}    A.~N.~Kolmogorov, Dokl.~Akad.~Nauk SSSR 31, 538 (1941).

\bibitem{Kraichnan}     R.~H.~Kraichnan, 
                        Physics of Fluids {\bf 10}, 1417 (1967).

\bibitem{Batchelor}     G.~K.~Batchelor, Physics of Fluids Supp.~II, 233 (1969).

\bibitem{Gotoh}         T.~Gotoh, Phys.~Rev. E {\bf 57}, 2984 (1998).

\bibitem{Chertkov}      M.~Chertkov, C.~Connaughton, I.~Kolokolov, and V.~Lebedev,
                        Phys.~Rev.~Lett. {\bf 99}, 084501  (2007).

\bibitem{Xia}           H.~Xia, D.~Byrne, G.~Falkovich, and M.~Shats,
                        Nat.~Phys. {\bf 7}, 321 (2011).

\bibitem{Brandenburg}   C.~Chan, D.~Mitra, and A.~Brandenburg, Phys.~Rev. E {\bf 85}, 036315 (2012).

\bibitem{Mininni}       P.~D.~Mininni and A.~Pouquet, Phys.~Rev. E {\bf 87}, 033002 (2013).

\end{thebibliography}
\end{document}